\documentclass[aip,jcp,reprint]{revtex4-1}

\usepackage[mathlines]{lineno} 
\usepackage{graphicx} 
\usepackage{amsmath} 

\makeatletter
\def\@email#1#2{%
 \endgroup
 \patchcmd{\titleblock@produce}
  {\frontmatter@RRAPformat}
  {\frontmatter@RRAPformat{\produce@RRAP{*#1\href{mailto:#2}{#2}}}\frontmatter@RRAPformat}
  {}{}
}%
\makeatother

\begin{document}


\preprint{AIP/123-QED}

\title{Locally tuned hydrodynamics of active polymer chains} 

\author{Lisa Sappl}
\affiliation{Faculty of Physics, University of Vienna, Boltzmanngasse 5, 1090 Vienna, Austria}
\affiliation{Vienna Doctoral School in Physics, University of Vienna, Boltzmanngasse 5, 1090 Vienna, Austria}

\author{Christos N. Likos}
\affiliation{Faculty of Physics, University of Vienna, Boltzmanngasse 5, 1090 Vienna, Austria}

\author{Andreas Z\"ottl}
\affiliation{Faculty of Physics, University of Vienna, Boltzmanngasse 5, 1090 Vienna, Austria}


\date{\today} 

\begin{abstract} 
We employ mesoscopic simulations to study active polymers in a solvent via multi-particle collision dynamics. We investigate linear chains in which either the head or tail monomer exerts an active force, directed away from or towards its neighbor, respectively, while the remaining monomers are passive. We find that, in contrast to flexible chains, for stiff chains the position of the active monomer has minimal influence on both the structural and dynamic properties of the chain. 
An active head monomer pulls the chain behind it, straightening the backbone -- an effect that can be interpreted as activity-induced stiffening. In contrast, an active tail pushes into the chain, causing crumpling. This leads to faster decorrelation of the polymer backbone over time, rendering the active motion less persistent. These effects occur regardless of whether hydrodynamic interactions are included or not. Hydrodynamics is included by the imposition of a local counter-force in the surrounding fluid, as opposed to distributing the former equally to all fluid elements. By specifying the position of this counterforce onto the fluid, we can tune the hydrodynamic flow fields of the active polymers being both contractile and extensile.
Interestingly, the emerging pusher- and puller flow fields are strongly influenced by the force propagation inside the polymer chain.
\end{abstract}


\maketitle 

\section{\label{sec:Intro}Introduction} 

In recent years, the field of \textit{active matter} has emerged as a promising line of research within the soft matter community, featuring new interesting physical phenomena. In general,
active matter is defined by its ability to make use of free energy either from within, or externally from its environment, to convert it into active motion such as active transport, rotation or enhanced diffusion.\cite{Vicsek2012, Ramaswamy2010}
Various biological systems are inherently out of equilibrium and feature non-equilibrium and active processes. Examples can be found at the macroscopic scale, where active agents are animals such as birds or sheep that move individually and interact within groups in such a way that their collective motion exhibits self-organizing patterns that are non-resembling of equilibrium states. At the microscopic scale, we find active microswimmers such as bacteria, parasites, or algae,\cite{Elgeti2015} and at the nano-scale cytoskeletal filaments, such as microtubules or actin that are propelled by motor proteins like kinesin and myosin.\cite{Mizuno2007, Yildiz2004} Noteworthy, the biopolymers DNA and RNA become active during processes such as replication and transcription by DNA- or RNA polymerase, respectively.\cite{Guthold1999, Mejia2015}

On the other hand, prominent examples of artificial active matter are active Janus colloids which are particles with half of their surface coated, e.g., by a material that locally catalyzes a chemical reaction in the surrounding fluid in such a way that it propels the particle as a consequence,\cite{Howse2007, Palacci2010} or locally heats up the colloid by self-thermophoresis.\cite{Jiang2010, Bregulla2014}
Artificial micro- and nanoswimmers also exhibit promise in the biomedical field, e.g., for the targeted delivery of drugs in health care applications.\cite{Wang2012, Bechinger2016}
From a theoretical and computational perspective, generic features of active matter such as motility-induced phase separation, boundary accumulation and enhanced transport have been investigated extensively.\cite{Cates2015, Bechinger2016, Zottl2023}

An important class of biologically relevant active matter are active polymers and filaments, displaying novel structural and dynamical behavior in contrast to their passive counterparts, as shown in various experimental and computational studies.
Different aspects of activity can be systematically analyzed using computational methods, and activity can be included at the monomer level.
In principle, monomer activity can be modeled either (i) as a scalar, i.e., an isotropic enhanced stochastic force displaying effectively as a locally enhanced temperature,\cite{Smrek2017,Smrek2020} or (ii) as a vector, i.e., a locally applied directed active force or velocity with some persistence.
The latter is often implemented in such a way that active monomers are modeled as
active Brownian particles, i.e., their active force or velocity points in a random direction and the direction changes diffusively over time.\cite{Loi2011, Kaiser2015, Bianco2018, Gomez2019, Das2021, Winkler2016, Siebert2017}
Compared to passive polymer conformations, these types of active polymers may swell or shrink.
Active polymers in which monomers have a preferred direction of activity are referred to as polar active polymers, and they exhibit vastly different phenomena in comparison to their active Brownian counterparts.\cite{Bianco2018} Examples include the observation of spontaneous spiral formation and break-up in polar active worm-like chains,\cite{Isele-Holder2015} conformational changes in ring polymers depending on polymer size,\cite{Locatelli2021} or studies on the effects of inertia \cite{Tejedor2024} or polydispersity \cite{Landi2025} on conformations and dynamics of polar active polymers. Most recently, polar active polymers in which only a selected few monomers are active have been studied and it was found that the density and location of the active sites significantly influence the conformational and dynamical properties of the macromolecules.\cite{Jaiswal2025} Active polymer models have further been helpful to understand collective phenomena of active living matter such as cyanobacteria \cite{Faluweki2023} and entangled worms.\cite{Nguyen2021,Sinaasappel2025} However, although most of the aforementioned systems occur in a fluid environment, the effect of hydrodynamic interactions (HI) had been largely ignored, despite the fact that HI can
lead to qualitatively different polymer dynamics.\cite{Winkler2020}

A prominent example of a computational method which captures both the intrinsic stochastic nature and hydrodynamic interactions naturally is the explicit, particle-based multi-particle collision dynamics (MPCD) method, sometimes also referred to as stochastic rotation dynamics (SRD).\cite{Malevanets1999, Gompper2009} MPCD is a mesoscale approach to model a fluid that helps bridge the time- and length scales necessary to simulate polymer systems,\cite{Gompper2009} microswimmers \cite{Zottl2020} and active fluids \cite{Kozhukhov2022, Baziei2025} while maintaining sufficient local information to capture hydrodynamic effects.
While previous work mainly focused on simulating rigid microswimmers,\cite{Downton2009, Gotze2010, Zottl2012, Zottl2018} bacteria \cite{Hu2015, Zottl2019, Martin2025, Wu-Zhang2025} or undulatory microorganisms \cite{Yang2008, Elgeti2010, Munch2016} in MPCD fluids, recently MPCD had been applied to model active polymers.
In contrast to \textit{dry} active matter, where active forces on monomers may not be balanced explicitly by forces on the surrounding, in \textit{wet} active matter, an explicit force-free setup, i.e., an active system without externally applied forces can be enforced.\cite{Ramaswamy2010,Marchetti2013}
Previous work includes simulations of active polymers built from active Brownian particles (ABPs) \cite{Llahi2022} without applying a counterforce to the fluid, active filaments in confinement with a counterforce applied to fluid particles within the multi-particle collision cells,\cite{Elgeti2009} active polymers with a tangential active force and a globally applied counterforce to the whole system,\cite{vanSteijn2024} and chemically active polymers.\cite{Jain2022,Jaiswal2024} We introduce in this work an alternative method, in which we locally apply a counterforce to the fluid, thus introducing a well-defined method to tune the local fluid velocity fields around active polymers whilst strictly maintaining an overall zero-force setup.

The rest of this work is organized as follows: Sec.~\ref{sec:methods} introduces the computational methods used, including the modeling of the polymer and the MPCD solvent. Sec.~\ref{sec:active_motion} shows results about the dynamic behavior, in particular the swimming velocity, and Sec.~\ref{sec:conformation} focuses on conformational properties of the active polymers. We show solvent density and flow fields in Sec.~\ref{sec:fields} and discuss the effect of hydrodynamic interactions in Sec.~\ref{sec:hydrodynamics}. Sec.~\ref{sec:monomer_resolved} offers a more zoomed-in understanding about active polymers by presenting results on a monomer resolved level, and Sec.~\ref{sec:force_dipole} discusses the internal force transmission and force dipole strength. Finally, in Sec.~\ref{sec:conclusions}, we summarize our findings and present further research suggestions. A short discussion on the special case
of active dimers is delegated to the Appendix.

\section{\label{sec:methods}Methods} 

We simulate a hybrid system in which we have both an active polymer and an explicit solvent present. In the following we introduce the polymer model. The solvent model, as well as polymer-solvent interaction will be discussed subsequently.

\subsection{Active MD polymer chain}

We model our active polymer using a bead-spring model, consisting of $N$ monomers, each of mass $M$, that interact with a Weeks-Chandler-Andersen (WCA) potential,\cite{Weeks1971}
%
\begin{align}
    U_{\mathrm{WCA}} (R_{ij}) = 
    \begin{cases}
        4\epsilon \left[ \left( \frac{\sigma}{R_{ij}} \right)^{12} - \left( \frac{\sigma}{R_{ij}} \right)^6 \right] &+ \epsilon,\\ & \text{if } 
        R_{ij} \leq 2^{1/6} \sigma, \\
        0, & \text{else} ,
    \end{cases}
\end{align}
between all monomer pairs, where $R_{ij} = |\boldsymbol{R}_i - \boldsymbol{R}_j|$ denotes the distance between monomers $i$ and $j$, with index $i,j \in \{ 1, ... , N \}$, and using their respective position vectors $\boldsymbol{R}_i$ and $\boldsymbol{R}_j$. Our length scale and energy scale are given by $\sigma$ and $\epsilon$ respectively. The unit of mass $m$ is given by the solvent particles (discussed later), thus giving the unit of time $\tau=\sqrt{m\sigma^2/\epsilon}$. Neighboring monomers are connected by adding the finitely extensible nonlinear elastic (FENE) potential,\cite{Kremer1990}
\begin{equation}
    U_{\mathrm{FENE}} (R_{i,i+1}) = -\frac{K}{2} R_{\mathrm{max}}^2 \ \log{ \left[ 1 - \left( \frac{R_{i,i+1}}{R_{\mathrm{max}}} \right)^2 \right] } ,
\end{equation}
for which we set the elasticity constant $K=30\,\epsilon/\sigma^2$, and maximum bond extension $R_{\mathrm{max}} = 1.5\sigma$. 
We additionally impose a bending potential \cite{Faller1999, Faller2000} 
$U_{\mathrm{bend}}(\theta_i)$ acting on the angle $\theta_i$ subtended between two 
consecutive bonds:
\begin{equation}
    U_{\mathrm{bend}} (\theta_i) = \kappa \left(1 + \cos \theta_i \right).
\end{equation}
Determination of the angle follows by the positions of the three consecutive monomers $\{i-1,i,i+1\}$ forming the two bonds that meet at angle $\theta_i$. The bending stiffness $\kappa$ defines resistance to bending and we will examine polymers of different stiffness by varying this parameter. The value for $\cos \theta_i$ at center monomer $i$ is given by the normalized scalar product of the two bond vectors,
\begin{equation}
    \cos \theta_i = \frac{(\boldsymbol{R}_{i+1}-\boldsymbol{R}_{i}) \cdot (\boldsymbol{R}_{i-1}-\boldsymbol{R}_{i})}{|\boldsymbol{R}_{i+1}-\boldsymbol{R}_{i}| |\boldsymbol{R}_{i-1}-\boldsymbol{R}_{i}|}
\label{eq:costhetai}
\end{equation}
one connecting monomer $i$ with monomer $i-1$ and the other connecting $i$ with $i+1$. 

Last, we add an active force to our model,
\begin{equation}
    \boldsymbol{F}_a = f_a \hat{\boldsymbol{e}}_f ,
\end{equation}
of magnitude $f_a$ and along a direction given by the unit vector $\hat{\boldsymbol{e}}_f$; the
force $\boldsymbol{F}_a$ is applied on the monomer with index $i_{\mathrm{act}}$. In this work, we model polymers in which only either the tail monomer or the head monomer is active while the remaining $N-1$ monomers remain passive. We denote the head monomer with the index $i=1$ and the tail monomer we denote with $i=N$. The direction of the active force
\begin{equation}
    \hat{\boldsymbol{e}}_f =
    \begin{cases}
    \frac{\boldsymbol{R}_1 - \boldsymbol{R}_2}{|\boldsymbol{R}_1 - \boldsymbol{R}_2|} & \text{if $i_{\mathrm{act}} = 1$,}\\
    \frac{\boldsymbol{R}_{N-1} - \boldsymbol{R}_N}{|\boldsymbol{R}_{N-1} - \boldsymbol{R}_N|} & \text{if $i_{\mathrm{act}} = N$,}
    \end{cases}
\end{equation}
aligns with the backbone of the polymer locally at the active monomer $i_{\mathrm{act}}$ (see also Fig.~\ref{fig:model}). If not stated otherwise, we set the magnitude of the active force $f_a=30\,\epsilon / \sigma$.

The monomer positions and velocities are propagated using the Velocity Verlet integration scheme,\cite{Verlet1967,Swope1982,Allen2017} in which the total force acting on monomer $i$ is given by the gradient of the total potential $U$ acting on it plus the active force via
\begin{equation}
    \boldsymbol{F}_{i} = - \boldsymbol{\nabla}_i U + \boldsymbol{F}_{a} \delta_{i,i_{\mathrm{act}}},
\end{equation}
using the Kronecker-Delta $\delta_{i,i_{\mathrm{act}}}$, and where
\begin{eqnarray}
    U = &&\sum_{i=1}^{N-1} \sum_{j=i+1}^{N} U_{\mathrm{WCA}} (R_{ij}) + \sum_{i=1}^{N-1} U_{\mathrm{FENE}} (R_{i,i+1}) \nonumber \\
    &&+ \sum_{i=2}^{N-1} U_{\mathrm{bend}} (\theta_i) .
\end{eqnarray}
%
We set the time step for the Velocity Verlet integration to $\delta t = 0.002\tau$.

\subsection{Coupling to an MPCD background fluid}

\begin{figure}
    \includegraphics[width=\linewidth]{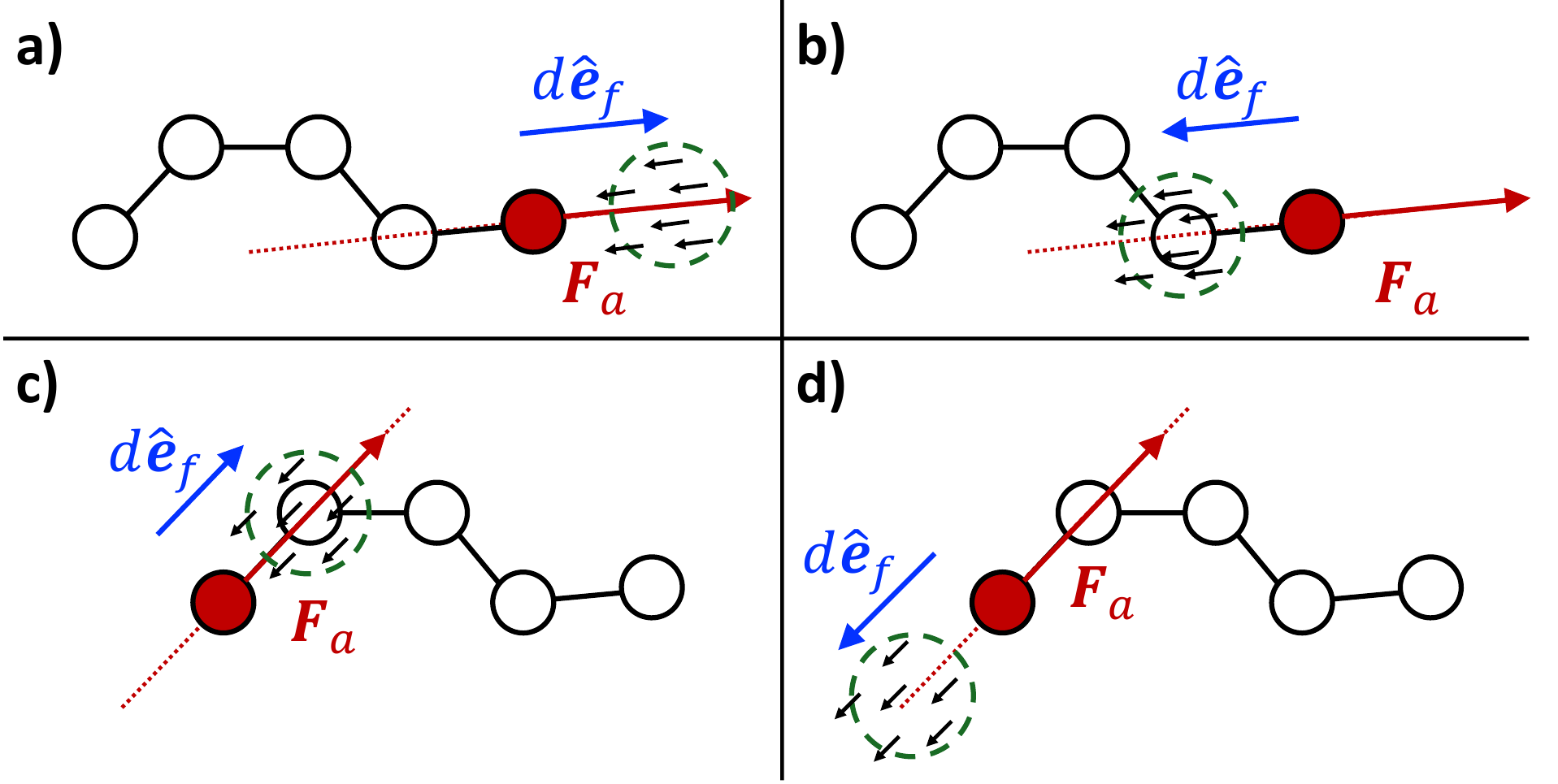}
    \caption{\label{fig:model}Schematic of our polymer that is a) head-active with counterforce placement before (hd$+$), b) head-active with counterforce placement behind (hd$-$), c) tail-active with counterforce placement before (td$+$), and d) tail-active with counterforce placement behind (td$-$) the active monomer. The active monomer is marked red, the counterforce volume is depicted by the green dashed circle. The blue arrow denotes the counterforce dispacement vector and the black arrows the distribution of the counterforce.}
\end{figure}

The active polymer is moving in a viscous fluid which we model explicitly using the MPCD method.\cite{Malevanets1999, Gompper2009} The MPCD solvent is a coarse-grained, particle-based representation of the fluid including both hydrodynamic interactions and thermal noise, 
built of point-like effective fluid particles of mass $m$ located at positions $\mathbf{r}_i(t)$ and moving at velocities $\mathbf{v}_i(t)$.
For the entire system (fluid+polymer) to be force-free, we need to apply a counterforce to the fluid particles that balances the active force applied to any active monomer. We do this by choosing a spherical volume with its center located at
\begin{equation}
    \boldsymbol{r}_{\mathrm{cf}} = \boldsymbol{R}_{i_{\mathrm{act}}} + d \hat{\boldsymbol{e}}_f ,
\end{equation}
and we define $d$ as our counterforce displacement. Concerning $d$, it can be tuned (discussed later), but for now we set it to either $+\sigma$ or $-\sigma$, therefore only distinguishing whether the counterforce center is displaced in a direction parallel to the active force vector (Fig.~\ref{fig:model}a,c), or anti-parallel to it (Fig.~\ref{fig:model}b,d). The counterforce itself is always anti-parallel to the active force.
As a result, we obtain four different models that we investigate in this work: the head-active polymer with counterforce displacement $d=+\sigma$, hereby called "hd$+$", the head-active polymer and $d=-\sigma$, hereby "hd$-$", and the two corresponding tail-active polymers "td$+$" and "td$-$". After determining the center of the counterforce volume, we determine the solvent particles within the spherical volume of diameter $2\sigma$, and distribute the counterforce among them evenly such that each will be given an equal fraction
\begin{equation}
    \boldsymbol{f}_{\mathrm{cf}} = - \boldsymbol{F}_a / n_{\mathrm{cf}} ,
\end{equation}
with $n_{\mathrm{cf}}$ representing the number of solvent particles within the counterforce volume. The quantity $n_{\mathrm{cf}}$ fluctuates, and while it is theoretically possible to find $n_{\mathrm{cf}} = 0$ for a given time step, given our simulation parameters, this is practically impossible to occur.

The dynamics of the fluid particles and coupling to the active polymer is realized by subsequent streaming and collision steps.
In the streaming step, the fluid particles outside the counterforce region move ballistically with time step $\delta t = 0.002 \tau$. For the fluid particles inside the counterforce region,
fluid particle positions and velocities are propagated following the Velocity Verlet algorithm
using the counterforce. 
After a well-defined number of streaming time steps $\delta t$, i.e., after a time $h$,
the solvent particles interact with each other as well as with the monomers in the collision step.
Therefore we divide the simulation box into cubic cells of side length $a$, and within each of these cells we determine the center-of-mass velocity
\begin{equation}
    \boldsymbol{v}_{\mathrm{cm}} = \frac{\sum\limits^{n_{\mathrm{c}}}_{i=1} m \boldsymbol{v}_i + \sum\limits^{N_{\mathrm{c}}}_{j=1} M \boldsymbol{V}_j}{m n_{\mathrm{c}} + M N_{\mathrm{c}}}, 
\end{equation}
including both the monomers (monomer velocities $\{\boldsymbol{V}_j\}$ and number of monomers in the cell $N_{\mathrm{c}}$) and solvent particles (solvent velocities $\{\boldsymbol{v}_i\}$ and number of solvent particles in the cell $n_{\mathrm{c}}$) within the cell.\cite{Malevanets2000b} Then, for each collision step and cell, a unit vector of random direction is drawn $\hat{\boldsymbol{n}}$, and all particles' relative velocities, i.e., velocities relative to the center-of-mass velocity, are rotated by a fixed angle $\alpha$ about the randomly drawn axis $\hat{\boldsymbol{n}}$, giving us the new velocity after rotation $\boldsymbol{v}_i'$ via
%
%
\begin{equation}
    \begin{cases}
    \boldsymbol{v}_i' (t) = \boldsymbol{v}_{\mathrm{cm}} (t) + \mathcal{D}(\hat{\boldsymbol{n}},\alpha) \left( \boldsymbol{v}_i (t) - \boldsymbol{v}_{\mathrm{cm}} (t) \right) , \\
    \boldsymbol{V}_j' (t) = \boldsymbol{v}_{\mathrm{cm}} (t) + \mathcal{D}(\hat{\boldsymbol{n}},\alpha) \left( \boldsymbol{V}_j (t) - \boldsymbol{v}_{\mathrm{cm}} (t) \right) , \\
    \end{cases}
    \label{eq:collision}
\end{equation}
of all solvent particles and monomers. 
The symbol $\mathcal{D}(\hat{\boldsymbol{n}},\alpha)$ stands for the rotation matrix corresponding to the axis $\hat{\boldsymbol{n}}$ and angle $\alpha$. This algorithm ensures that momentum is exchanged between monomers and solvent particles, but total local momentum is conserved, therefore correctly reproducing hydrodynamic interactions.\cite{Malevanets1999} To ensure Galilean invariance, the cell grid needs to be shifted by a random vector, components of which are drawn from a uniform distribution $U\sim[0,a)$ for each time step in which the stochastic rotation is performed.\cite{Ihle2001} After collision, we apply thermostatting to the monomers and solvent particles using cell-level Maxwell-Boltzmann scaling.\cite{Huang2010}
Employing this algorithm, total angular momentum is not conserved but was observed to fluctuate around zero. Furthermore, the system is not strictly torque-free as the distribution of solvent particles within the counterforce volume fluctuates. However, also the torque fluctuates around zero, preventing build up angular momentum.

For the polymer we set $N=10$ and $M=20m$. We use a simulation box of side length $L=20\sigma$, set the collision angle to $\alpha=2\pi/3$, the side length of the collision cells to $a=\sigma$, and operate at temperature $k_{\mathrm{B}}T=\epsilon$, all if not explicitly stated otherwise.
The solvent particle density $\rho_{\mathrm{sol}}=20\,\sigma^{-3}$ and collision time step $h=0.01\,\tau$ give us a relatively large dynamic viscosity $\eta = 158.4\, m/(\sigma\tau)$ of the MPCD solvent.\cite{Kikuchi2003}

%
%
%

\section{\label{sec:active_motion}Active motion} 

We first equilibrate passive polymers and
thereafter,
we reach steady-state active polymer conformations via a short pre-simulation of our systems over a time span of $\Delta t_{\mathrm{pre}} = 1000\,\tau$. 
Subsequent runs are performed over a time span of $\Delta t_{\mathrm{run}} = 10000\,\tau$, during which measurements are taken. Still, we disregard the first $\Delta t_{\times} = 1000\,\tau$ of the run for solvent start-up purposes.
We also simulate a passive system ($f_a=0$) with the same parameters to compare it to the active systems.

We find that our active polymers swim, i.e., they perform persistent random motion, as opposed to the purely Brownian motion that the passive polymer exhibits.
We quantify this persistent random motion via the swimming velocity of the polymer $v_0$, which we define as the center-of-mass velocity along the polymer backbone represented by the end-to-end unit vector
\begin{equation}
    \hat{\boldsymbol{e}}_{\mathrm{ee}} = \frac{\boldsymbol{R}_1 - \boldsymbol{R}_N}{|\boldsymbol{R}_1 - \boldsymbol{R}_N|} .
\end{equation}
We determine the center-of-mass velocity via the center-of-mass displacement over time, and thus 
calculate the swimming velocity at time $t$ as
\begin{equation}
    v_0(t) = \frac{\boldsymbol{R}_{\mathrm{cm}}(t+\Delta t) - \boldsymbol{R}_{\mathrm{cm}}(t)}{\Delta t} \cdot \frac{\hat{\boldsymbol{e}}_{\mathrm{ee}} (t+\Delta t) + \hat{\boldsymbol{e}}_{\mathrm{ee}} (t)}{2},
\end{equation}
%
using the center-of-mass position vector defined as $\boldsymbol{R}_{\mathrm{cm}}(t) = \sum_i \boldsymbol{R}_i(t) / N$, evaluated every $\Delta t = 1\,\tau$.

\begin{figure*}
    \includegraphics[width=\linewidth]{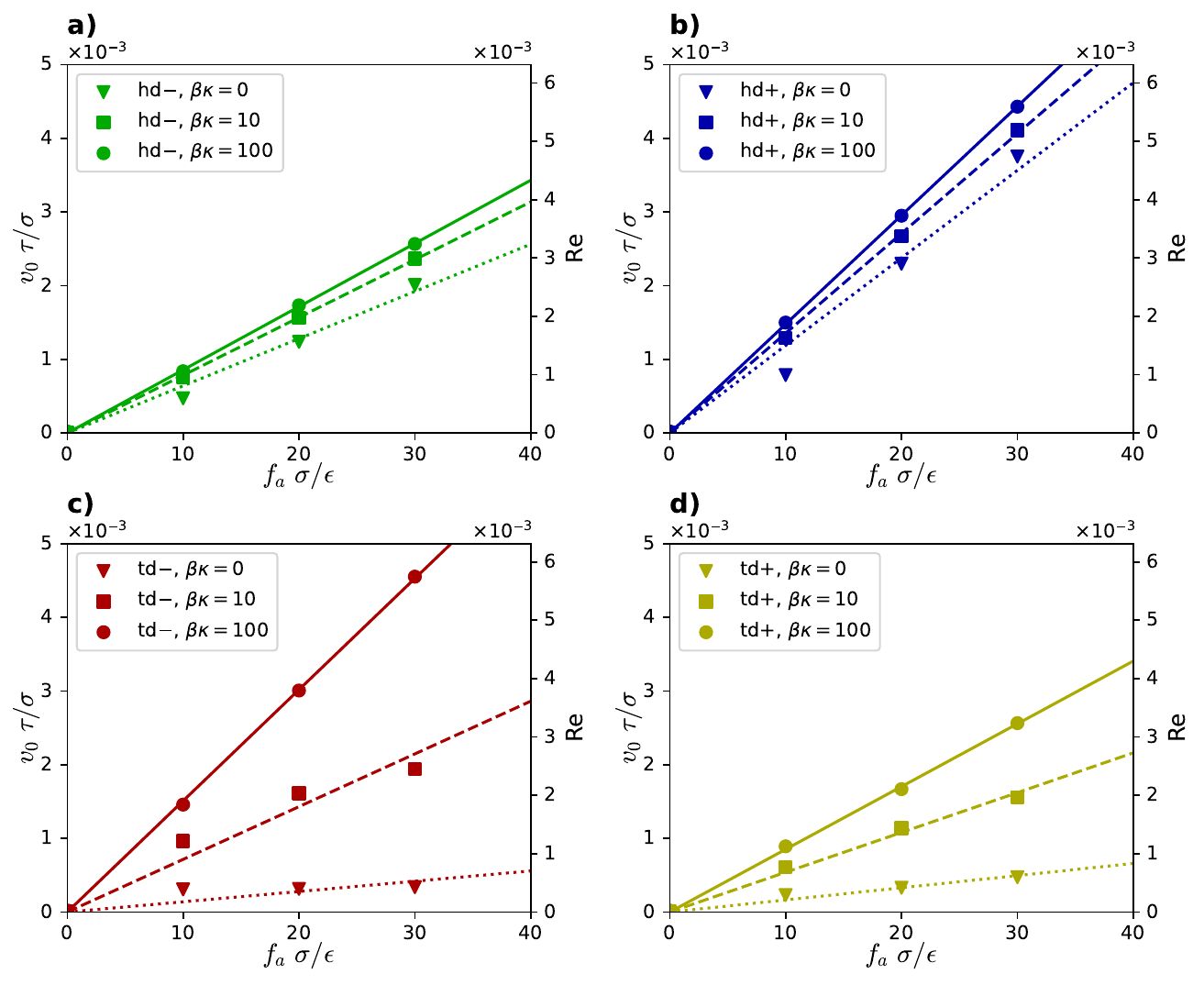}
    \caption{\label{fig:v0_fa}Swimming velocity $v_0$ against the active force $f_a$ for a) the hd$-$ model, b) the hd$+$ model, c) the td$-$ model, and d) the td$+$ model, for varying stiffness parameters $\beta\kappa$ ($\beta = 1/k_{\mathrm{B}}T$). We calculate $v_0$ as a time- and ensemble average over a time span of $\Delta t_{\mathrm{span}} = 9000\,\tau$ and 48 samples. Error bars are smaller than symbol size. The lines represent linear fits forced through (0,0).}
\end{figure*}

Since we aim at modeling active polymers at the nano- or micrometer scale, they operate at very low Reynolds numbers $\mathrm{Re} \ll 1$ where inertia can be neglected entirely.\cite{Happel2012,KimKarila}
To ascertain that this is indeed the case and to quantify the relevance of the effects of inertia, we calculate the resulting Reynolds number in our system
\begin{equation}
    \mathrm{Re} = \frac{l_{\mathrm{c}} m \rho_{\mathrm{sol}}}{\eta} v_0,
\end{equation}
choosing the contour length $l_{\mathrm{c}}=N\sigma$ as a characteristic size of the swimmer. Indeed, we find that $\mathrm{Re} \lesssim 6\times 10^{-3}$.
As expected in the low Reynolds number regime, we obtain a linear relationship between swimming velocity $v_0$ and active force magnitude $f_a$ for (sufficiently) stiff rodlike active polymers (Fig.~\ref{fig:v0_fa}).
However, for more flexible polymers we observe a non-trivial relationship between $v_0$ and $f_a$.
In general, the swimming velocities are smaller when the polymers are more flexible.
Moreover, the head-active swimmer is much less sensitive to changes in stiffness parameter $\kappa$ than the tail-active swimmer, both effects motivating deeper investigation.

\begin{figure}
    \includegraphics[width=\linewidth]{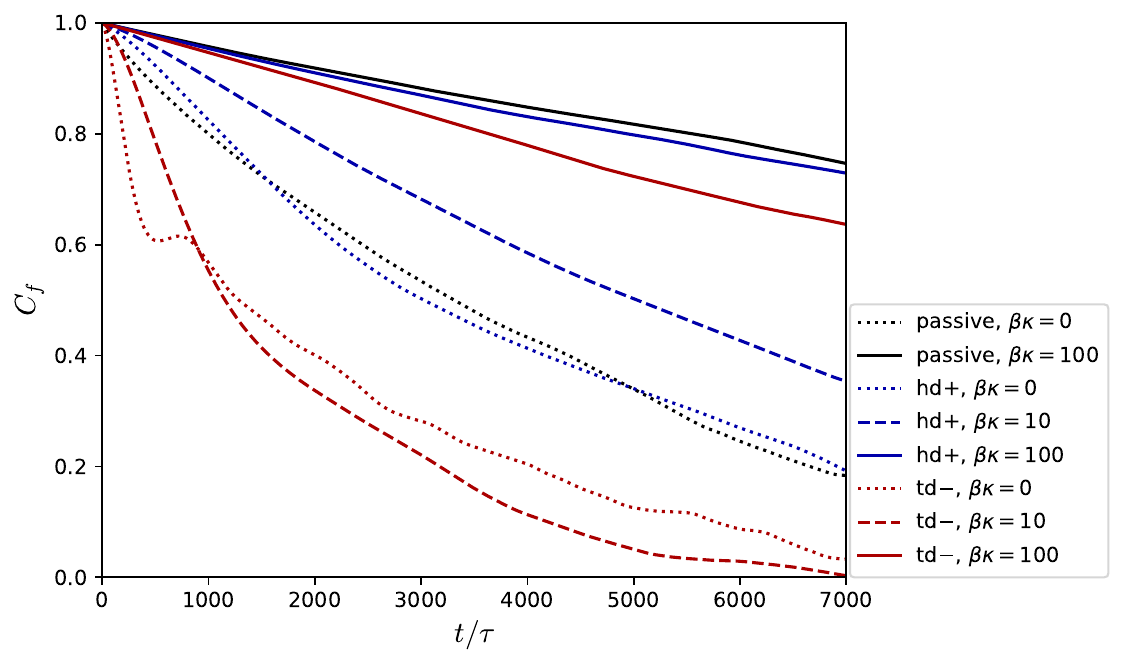}
    \caption{\label{fig:acf_eee}Auto-correlation function $C_f$ of the end-to-end unit vector $\hat{\boldsymbol{e}}_{ee}$ in different polymer models over time ($\beta = 1/k_{\mathrm{B}}T$). The curves show averages over 48 runs, $f_a=30\,\epsilon/\sigma$.}
\end{figure}

We further investigate the auto-correlation-function of the polymer backbone orientation in the steady-state over time, calculated following
\begin{equation}
    C_f(t) = \left\langle \frac{\hat{\boldsymbol{e}}_{ee}(t_0 + t) \cdot \hat{\boldsymbol{e}}_{ee}(t_0) }{\hat{\boldsymbol{e}}_{ee}(t_0) \cdot \hat{\boldsymbol{e}}_{ee}(t_0)} \right\rangle .
\end{equation}
%
We expect that passive polymers exhibit an exponential decay in their $C_f$ over time,\cite{Rubinstein2003}
\begin{equation}
    C_f(t) \sim e^{-t/\tau_r}
\end{equation}
with a slower decay for stiffer chains, where $\tau_r$ is the orientational correlation time, in agreement with our findings (Fig.~\ref{fig:acf_eee}). 
Regarding the active polymers, $\tau_r$ varies among the different implementations, i.e.\
$\tau_r \approx 2 \times 10^3\tau - 2 \times 10^4\tau$.
However, while in good approximation, the orientation-correlation of head-active polymers exhibit an exponential decay,
tail-active polymers show exponential decay only for the stiff polymer. The fully flexible tail-active polymer even exhibits some kind of periodicity at the time scale of $t \sim 10^3\tau$. In Fig.~\ref{fig:acf_eee} and subsequent figures, only the hd$+$ and td$-$ active polymer models are represented. This is because aside from differences in swimming velocity $v_0$, the hd$-$ model behaves almost identically to the hd$+$ model, and the 
td$+$ model is almost identical to td$-$ one, concerning all effects discussed hereafter.

An analysis of the mean-squared displacement of the polymer center-of-mass over time, $\langle (\Delta \boldsymbol{R}_{\mathrm{cm}})^2 (t) \rangle$, shows a similar trend (Fig.~\ref{fig:msd_poly}):
\begin{figure*}
    \includegraphics[width=\linewidth]{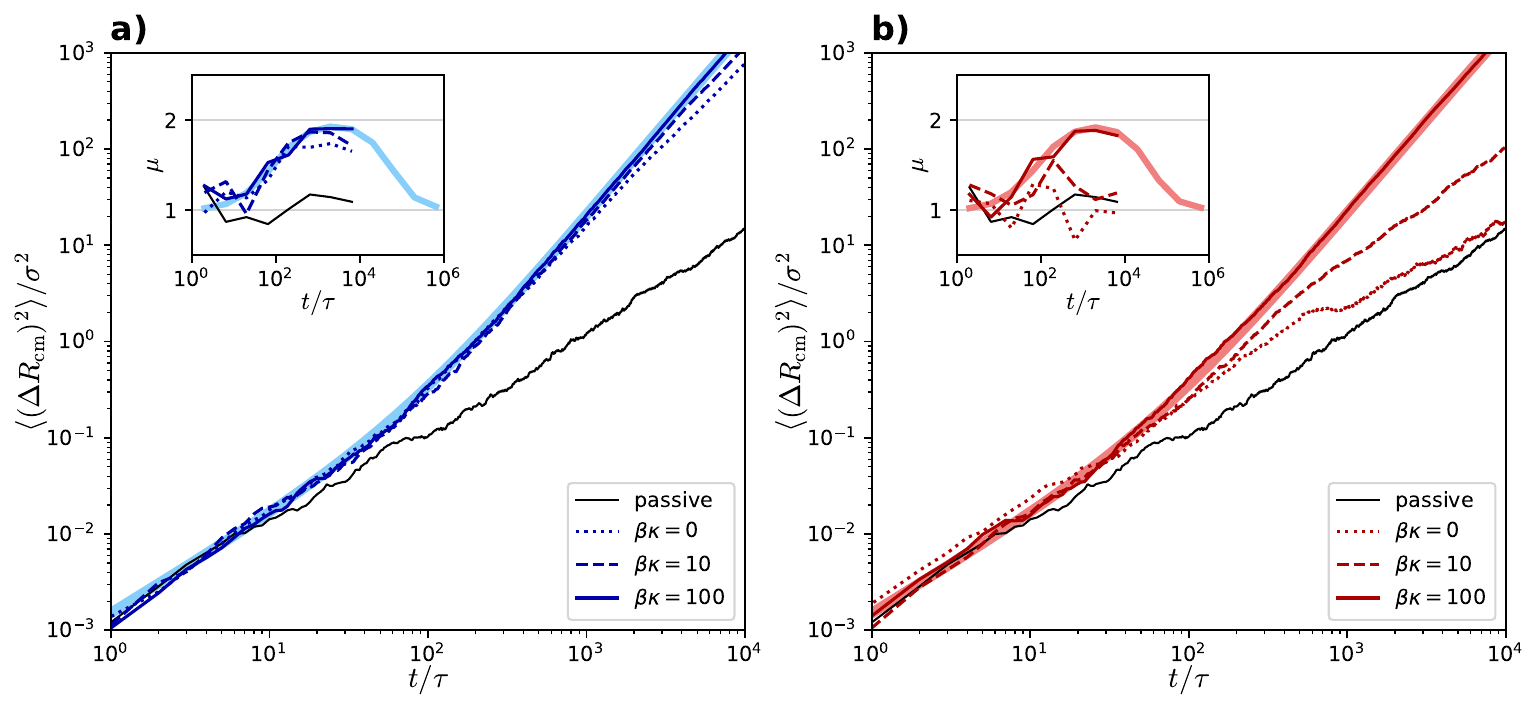}
    \caption{\label{fig:msd_poly}Mean squared displacement (MSD) $\langle (\Delta \boldsymbol{R}_{\mathrm{cm}})^2 (t) \rangle$ over time for a) the hd$+$ polymer, and b) the td$-$ polymer. The insets show the instantaneous exponents $\mu$ of the scaling $\langle (\Delta R_{\mathrm{cm}})^2 \rangle \propto t^{\mu}$ of the corresponding mean-squared displacement curves. The bright bold curves represent the theoretical MSDs of persistent random walks for stiff particles following Eq.~(\ref{eq:persistent_rw}). Note the different extent of the time domain in the insets. Results pertain to an active force $f_a=30\, \epsilon/\sigma$.}
\end{figure*}
The passive polymer displays diffusive motion across all time scales, independent of the stiffness parameter $\kappa$. The head-active polymer exhibits diffusive motion for very short time scales, and transitions to the ballistic regime at $t \sim 10^2\,\tau$. This is attributed to the directed active motion in this time scale, as the backbone orientation vector decorrelates at larger time scales than the time scale shown for all three parameters of stiffness. Regarding the tail-active polymer, we again find that only the stiff polymer fully enters the ballistic regime. The semi-flexible polymer's orientation decorrelates at roughly the same time scale as the ballistic regime would manifest itself. This gives enhanced diffusion for a small time window, but returns to diffusive motion soon after. The fully flexible tail-active polymer even shows a short sub-diffusive bump right after a short super-diffusive onset. This bump correlates with the time scale of the bump in the auto-correlation function and stems from a "cat's tail"-like motion (see supplementary videos S1.mov, S2.mov, and S3.mov). 

We can compare the mean-squared displacement curves of the active polymers with the persistent random walk model for stiff axisymmetric active particles,\cite{Zottl2016} given by the the expression: 
\begin{equation}
    \langle (\Delta \boldsymbol{R}_{\mathrm{cm}})^2 (t) \rangle = 6 D_{\mathrm{pass}} t + 2 v_0^2 \tau_r t - 2 v_0^2 \tau_r^2 (1 - e^{-t/\tau_r}).
    \label{eq:persistent_rw}
\end{equation}
Here, $D_{\mathrm{pass}}$ is the diffusion coefficient of the corresponding passive polymer,
which we obtain as $D_{\mathrm{pass}} \approx 2 \times 10^{-4} \sigma^2/\tau$ by fitting the 
long-time mean-squared displacement according to 
\begin{equation}
    \langle (\Delta \boldsymbol{R}_{\mathrm{cm, pass}})^2 (t) \rangle = 6 D_{\mathrm{pass}} t .
\end{equation}
Using the measured values of $D_{\mathrm{pass}}$, $\tau_r$ and $v_0$ allows us to quantify the importance of active vs.\ diffusive motion. Both the P\'eclet number
\begin{equation}
    \mathrm{Pe} = \frac{v_0 l_{\mathrm{c}}}{D_{\mathrm{pass}}} \sim 10^1 - 10^2
\end{equation}
as well as the persistence number \cite{Zottl2016}
\begin{equation}
    \mathrm{Pe}_r = \frac{v_0 \tau_r}{l_{\mathrm{c}}} \sim 10^0 - 10^1
\end{equation}
are larger than unity, confirming the importance of activity.
Indeed the persistent random walk model from Eq.~(\ref{eq:persistent_rw}) provides an excellent prediction for the mean-squared displacement curves of the stiff active polymers shown in Fig.~\ref{fig:msd_poly}. It lets us estimate the time scale at which the polymer enters the second diffusive regime, which begins at $t_{\mathrm{diff}} \approx 10^6\,\tau$ and will persist from there on. This time scale is not accessible with our simulation parameters and computational capacity. Therefore, in the insets of Fig.~\ref{fig:msd_poly}, at long times, only the calculated persistent random walk curves are shown, and no simulation data is available. However, to convince the reader that the active polymer does indeed enter a second diffusive regime, we repeated the simulations using dimers ($N=2$) that have much smaller $\tau_r$, and need a much smaller simulation box, therefore giving us access to the relevant time scale (see Appendix~\ref{sec:dimers}).

\section{\label{sec:conformation}Conformational properties} 

The presence of activity has ramifications not only for the polymer dynamics, but also for the conformational properties. The first parameter that is related to polymer size, is the radius of gyration $R_{\mathrm{G}}$ given by \cite{Rubinstein2003}
\begin{equation}
    R_{\mathrm{G}}^2 := \frac{1}{N} \sum\limits^N_{j=1} \left| \boldsymbol{R}_j - \boldsymbol{R}_{\mathrm{cm}} \right|^2 .
\end{equation}
Results for the radius of gyration $R_{\mathrm{G}}$ of the different polymers are given by Tab.~\ref{tab:rgyr}.
\begin{table}[h]
\caption{\label{tab:rgyr}Radii of gyration $R_{\mathrm{G}}$, in units of $\sigma$, for each polymer type, calculated from the squared radius of gyration averaged over a time frame of $\Delta t_{\mathrm{span}} = 9000\,\tau$ and 48 samples ($\beta = 1/k_{\mathrm{B}}T$, $f_a=30\,\epsilon/\sigma$).}
\begin{ruledtabular}
\begin{tabular}{lccc}
     & $\beta\kappa=100$ & $\beta\kappa=10$ & $\beta\kappa=0$ \\
    \hline
    passive & 2.76 $\pm$ 0.00 & 2.55 $\pm$ 0.01 & 1.65 $\pm$ 0.01 \\
    hd$-$ & 2.79 $\pm$ 0.00 & 2.66 $\pm$ 0.00 & 2.27 $\pm$ 0.01 \\
    hd$+$ & 2.80 $\pm$ 0.00 & 2.69 $\pm$ 0.00 & 2.44 $\pm$ 0.01 \\
    td$-$ & 2.73 $\pm$ 0.00 & 2.30 $\pm$ 0.01 & 1.59 $\pm$ 0.01 \\
    td$+$ & 2.73 $\pm$ 0.00 & 2.39 $\pm$ 0.01 & 1.57 $\pm$ 0.01 \\
\end{tabular}
\end{ruledtabular}
\end{table}
While stiff polymer chains are similar in size, with very little influence from the site of activity (head vs. tail), flexible polymer chains are slightly straightened in the case of head-monomer activity, or crumpled in the case of tail-monomer activity. Again, the site of activity is the most significant contributor to conformational effects, the location of the counterforce is less important. Hence, we will focus on the hd$+$ and td$-$ models in the following as representative situations. 

In Eq.~(\ref{eq:costhetai}), we defined the 
local angle $\theta_i$, $i = 2,3,\ldots,N-1$, formed between two successive
bonds sharing an apex at the monomer $i$. After 
constructing $N - 2$ histograms of the 
orientational cosines of the angles $\theta_i$
and obtaining thereby 
the probability densities $p(\cos\theta_i)$, 
we perform one final average over those as:
\begin{equation} 
    p(\cos{\theta}) = \frac{1}{N - 2}\sum_{i = 2}^{N-1}p(\cos{\theta_i}),
    \label{eq:costheta}
\end{equation}
yielding a probability density $p(\cos\theta)$.
Note that the quantity $\cos\theta$ appearing
in Eq.~(\ref{eq:costheta}) above is not the 
cosine of any physical angle $\theta$ in the
problem; rather, $\cos\theta$ is just a
parametrization for a variable
whose magnitude is bounded by unity; accordingly,
$\theta$ can be understood as an average bond
angle only in a loose way when it comes to the 
active chains. However, the probability
density $p(\cos\theta)$ serves as a tool for
mapping the active polymers onto passive ones with
an effective, activity-modifed bending rigidity
$\tilde\kappa(\kappa; {\mathrm{Pe}}, {\mathcal A})$, 
where ${\mathcal{A}} \in \{\mathrm{hd+}, \mathrm{hd-},\mathrm{td+},\mathrm{td-} \}$,
as we demonstrate below. 
\begin{figure*}
    \includegraphics[width=\linewidth]{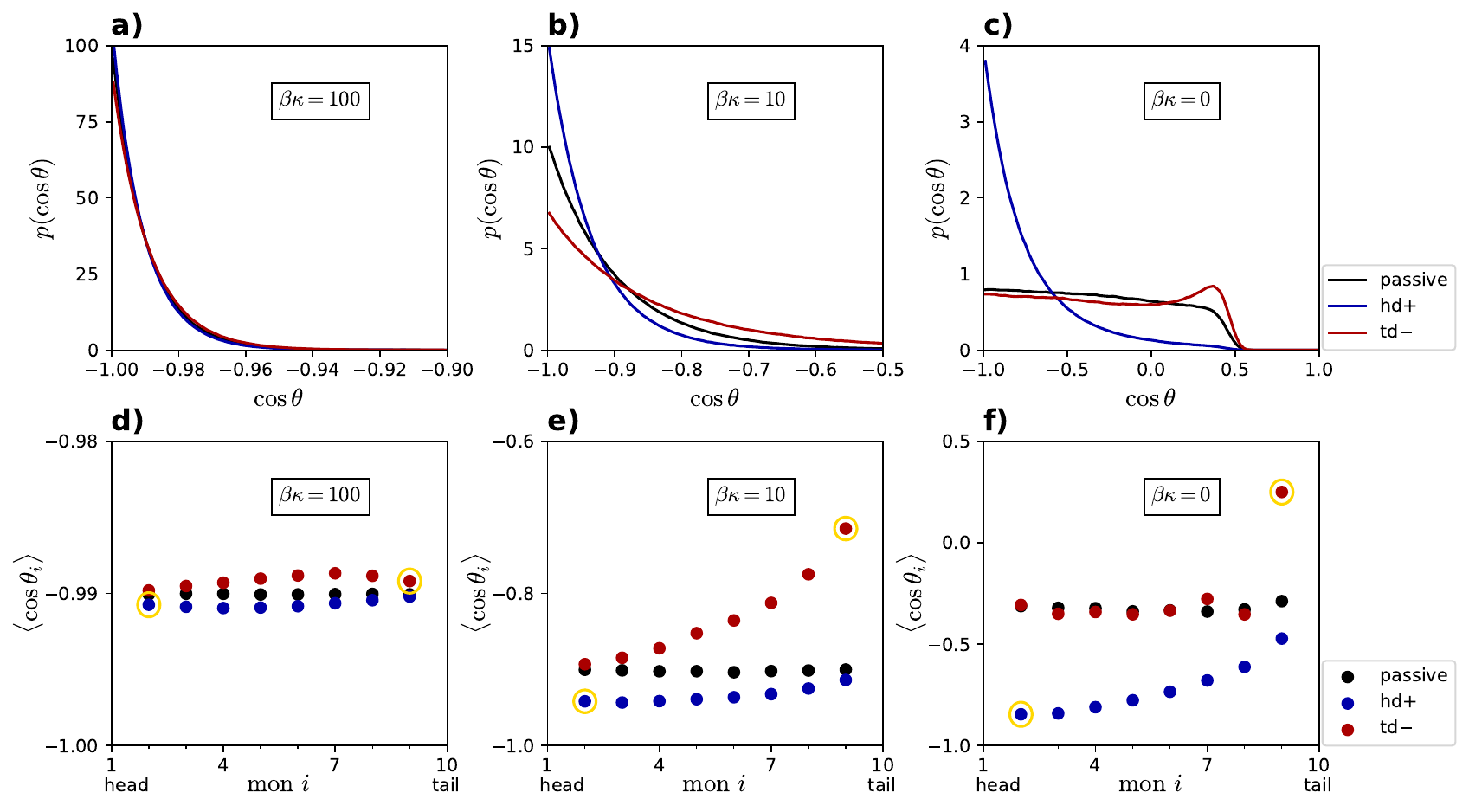}
    \caption{\label{fig:cos_theta}a-c) Probability density function of the variable $\cos{\theta}$, as defined in 
    Eq.~(\ref{eq:costheta}), for a) the stiff ($\beta\kappa=100$), b) the semi-flexible ($\beta\kappa=10$), and c) the fully flexible ($\beta\kappa=0$) polymer types. The theoretical distribution following Eq.~(\ref{eq:cos_theta_dist}) using proper normalization agrees very well with the passive polymer line in a) and b) (not shown). d-f) Average $\cos{\theta}$ per monomer site $i$ in d) the stiff ($\beta\kappa=100$), e) the semi-flexible ($\beta\kappa=10$), and f) the fully flexible ($\beta\kappa=0$) polymer types. The angle involving the active site is marked by a yellow circle for better readability (all plots shown for $f_a=30\, \epsilon/\sigma$).}
\end{figure*}

For the stiff polymers (Fig.~\ref{fig:cos_theta}a), the bending potential is very strong and therefore the active forces hardly influence the polymer conformation compared to a passive polymer. For high stiffness, $\beta\kappa \gg 1$, excluded volume interactions can be neglected as they only play a role for neighboring monomers to create the spring potential, and the $\cos{\theta}$ are distributed following an exponential law with
an effective, activity-modified bending stiffness
$\tilde\kappa$, i.e.:
\begin{equation}
    p(\cos{\theta}) \propto e^{-\beta U_{\mathrm{bend}}(\theta)} \propto e^{-\beta \tilde\kappa\cos{\theta}} .
    \label{eq:cos_theta_dist}
\end{equation}
This is still a good enough approximation for semi-flexible polymers with $\beta\kappa=10$ (Fig.~\ref{fig:cos_theta}b), although now the site of activity changes the beding rigidity, making it appear stiffer ($\tilde\kappa > \kappa$) in case of the head-active, or more flexible ($\tilde\kappa < \kappa$) in case of the tail-active polymer. In an ideal chain, the $\cos{\theta}$ are distributed uniformly. In a real chain, excluded volume effects make the phase space of the higher $\cos{\theta}$-region inaccessible, the limit posing $\cos{\theta} \approx 0.5$ which corresponds to the monomer triplet forming an equilateral triangle. Fully flexible ($\kappa = 0$) head-active polymers assume conformations that resemble polymers of finite stiffness,
$\tilde\kappa \ne 0$, Fig.~\ref{fig:cos_theta}c. On the other hand, the fully flexible tail-active polymers more closely follow the uniform-like $\cos{\theta}$-distribution of their passive counterparts up until $\cos{\theta} \approx 0$, then probabilities for small $\theta$ start to increase and show a maximum shortly before $\cos{\theta} \approx 0.5$, from where on the probability density assumes the value zero due to excluded volume interaction.

It is pertinent to pose the question whether these probability density functions are valid across all sites or whether effects originating from active monomers are screened over the contour length. To address it, we evaluate the expectation value 
$\langle\cos{\theta_i}\rangle$ of the cosine
of the angle $\theta_i$ specific to each monomer site for all polymer types, see Figs.~\ref{fig:cos_theta}d-\ref{fig:cos_theta}f. As it turns out, the angle that is closest to the active monomer, is in general also most strongly affected by the activity, and therefore it deviates the most from the equilibrium average angle in the passive polymer. As long as there is some degree of stiffness, this effect can be observed along the whole polymer chain, and albeit weakening with contour distance from the active site, it is never fully screened. In the fully flexible case, this is still true for head-active polymers, which experience effective stiffening. In the tail-active polymers, only the angles between bonds immediately adjacent to the active site are affected, due to a local crumpling of the polymer there. After that, the effect of activity is mostly screened along the contour length.

So far, only the location of the active site within the polymer exhibited drastic differences in polymer dynamics and conformational properties. The location of the counterforce volume seems to play a minor role, though. The hd$+$ polymer swims faster than the hd$-$ polymer, and the td$-$ polymer (given it is sufficiently stiff) swims faster than the td$+$ polymer. This can be explained by the placement of the counterforce volume being within the polymer region (hd$-$ and td$+$), slowing the polymer down following monomer-solvent interaction, or outside of the polymer region (hd$+$ and td$-$). Other than the magnitude of swimming speed, the placement of the counterforce volume has little effect on the active polymer conformations.

\section{\label{sec:fields}Hydrodynamic flow fields} 

For the hd$-$ and td$-$ models, we apply an extensile force dipole between the active monomer and the nearby solvent, i.e., the two force vectors are displaced such that they point away from each other. As a result one could naively expect a pusher flow field to emerge,\cite{Lauga2009,Zottl2016}
where fluid flows away from the swimmer along the dipole axis and towards the swimmer perpendicular to it. Analogously, for the hd$+$ and td$+$ models, we apply a contractile force dipole, i.e., the two force vectors point towards each other. Here, one would naively expect a resulting puller flow field, meaning that fluid flow is directed inwards along the dipole axis, and away from the swimmer perpendicular to it. 

\begin{figure*}
    \includegraphics[width=\linewidth]{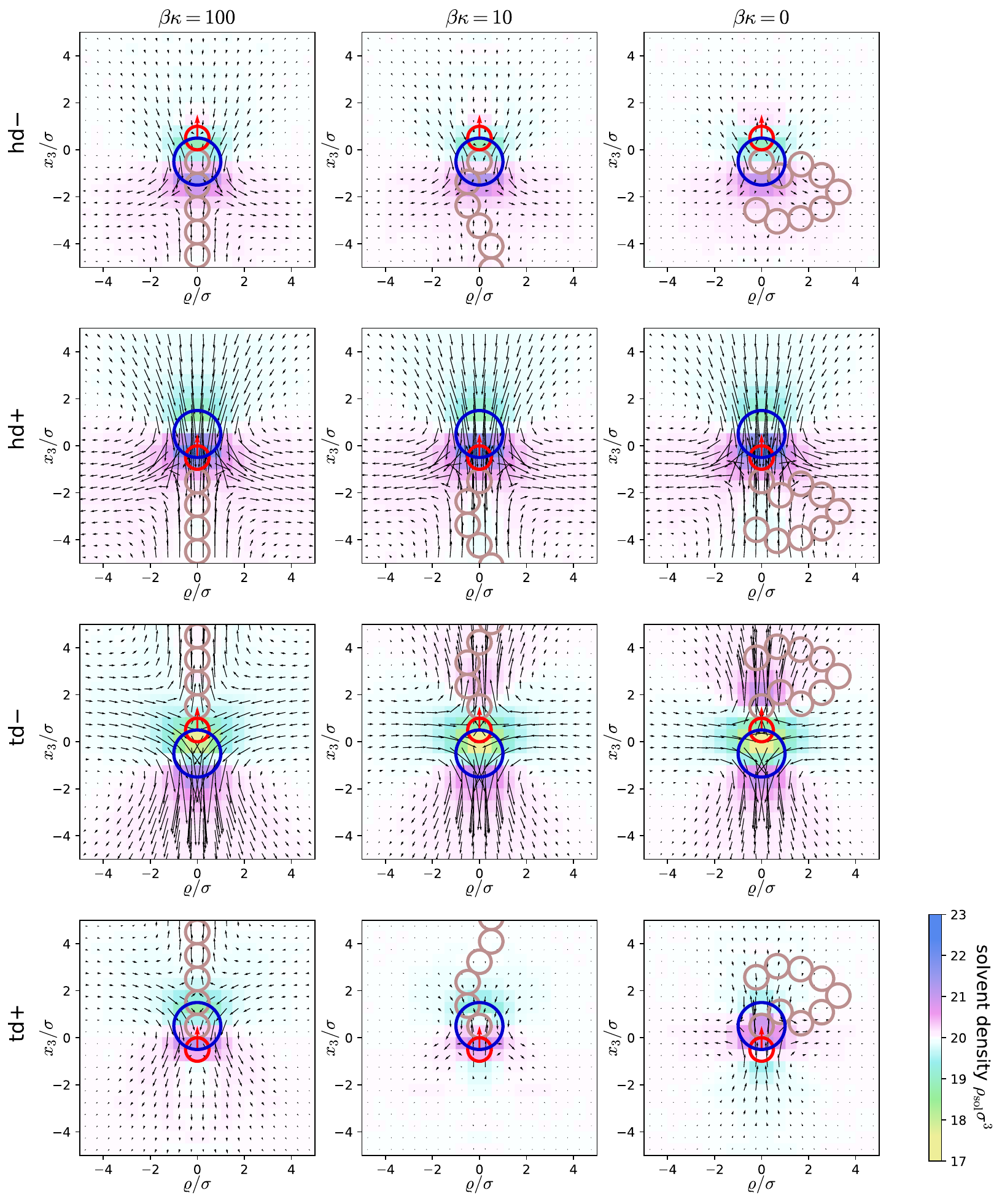}
    \caption{\label{fig:flow_fields}Flow fields for different types of active polymers for an active force strength $f_a=30\, \epsilon/\sigma$. A schematic representation of the polymer is depicted by the bright and light red circles, indicating monomers. The bright red circle represents the active monomer, and the bright red arrow its active force vector. The dark blue circle indicates the region in which the counterforce is exhibited on solvent particles.}
\end{figure*}

We calculate the fluid flow fields by transforming the solvent positions and velocities in each time step, such that the coordinate system center aligns with the force dipole center, and the $x_3$-axis aligns with the active force vector. Then we use cylindrical coordinates $\{\varrho,\varphi,x_3\}$ and
we average both solvent densities and velocity fields over $\varphi$ within a small region of $\Delta x_3 \times \Delta \varrho = 0.5a \times 0.5a$
and determine the fluid density and velocity field along the unit vectors $\hat{\boldsymbol{e}}_{\varrho}$ and $\hat{\boldsymbol{e}}_3$. Fig.~\ref{fig:flow_fields} illustrates the local solvent density and flow fields for different types of active polymers.

Because of the inherent compressibility of the MPCD solvent consisting of point-like fluid particles, the fluid density is not perfectly constant as for incompressible fluids; instead, density inhomogeneities caused by the applied active forces arise. Because of the finite speed of sound, the applied counterforce induces a solvent flow leading to a small depletion behind and small accumulation in front of it. In our  system these inhomogeneities are relatively small but are significant in particular close to the applied force dipole with $\rho_{\mathrm{sol}} \pm \sim 10\%$, as indicated in Fig.~\ref{fig:flow_fields} by the color bar.

 Concerning the solvent velocities, we find that the hd$+$ models exhibit puller flow fields, and the td$-$ models exhibit pusher flow fields for all degrees of stiffness $\kappa$ as expected. However, the stiff hd$-$ and stiff td$+$ models exhibit puller- and pusher flow fields respectively as well. At first glance, this is surprising, as we would expect puller flow fields for the contractile force dipoles ($d=+\sigma$), and pusher flow fields for the extensile force dipoles ($d=-\sigma$). Moreover, the more flexible td$+$ polymers seem to transform to exhibiting the naively expected puller flow fields with decreasing stiffness, whereas the hd$-$ flow fields remain puller types. Hence, despite applying a simple force dipole locally, the resulting forces in the system are more complex: The active force applied to one monomer is transmitted to other monomers through the polymer bonds, where also higher order multipoles are present.
In fact, in the case of a perfectly stiff polymer rod, the force on the fluid will be the center of the rod independently of the monomer 
on which the active force is applied, because the latter is instantaneously transmitted to the other monomers.
As a consequence, the stiff hd$-$ and td$+$ models, 
experience puller- and pusher flow fields, respectively.
Furthermore, this explains why the more flexible polymers experience a different flow field than the stiff polymers, especially for the td$+$ case.

\section{\label{sec:hydrodynamics}Effects of hydrodynamic interaction} 

To investigate which of the observed effects are due to hydrodynamic interactions (HI), and which are solely a consequence of polymer activity, we repeat our simulations using a modified version of the MPCD algorithm where hydrodynamic interactions are turned off while keeping all other fluid properties (i.e., viscosity and fluctuations) unchanged. Instead of simulating explicit solvent particles to which the counterforce is applied, we let the monomers exchange momentum with a random MPCD solvent acting as a simple thermostat.\cite{Ripoll2007} For each monomer $i$, we draw an effective solvent momentum $\boldsymbol{P}_i$ from a Maxwell-Boltzmann distribution of variance $m\tilde{\rho}_{\mathrm{sol}}k_{\mathrm{B}}T$ to interact with. The center-of-mass velocity in the hypothetical cell for each monomer $i$ is then given by
\begin{equation}
    \boldsymbol{v}_{\mathrm{cm},i} = \frac{M\boldsymbol{V}_i + \boldsymbol{P}_i}{M+m\tilde{\rho}_{\mathrm{sol}}} ,
\end{equation}
and we draw the local solvent density $\tilde{\rho}_{\mathrm{sol},i}$ from a Poisson distribution with expectation value $\lambda = \rho_{\mathrm{sol}}$ to respect density fluctuations as well. The new monomer velocity $\boldsymbol{V}_i'$ is calculated as usual following Eq.~(\ref{eq:collision}).
Note, since in this approach there is no counterforce to the fluid, the total force in the system is non-zero, similar to conventional approaches of modeling \textit{dry} active polymers neglecting hydrodynamic interactions.\cite{Winkler2020}

\begin{figure*}
    \includegraphics[width=\linewidth]{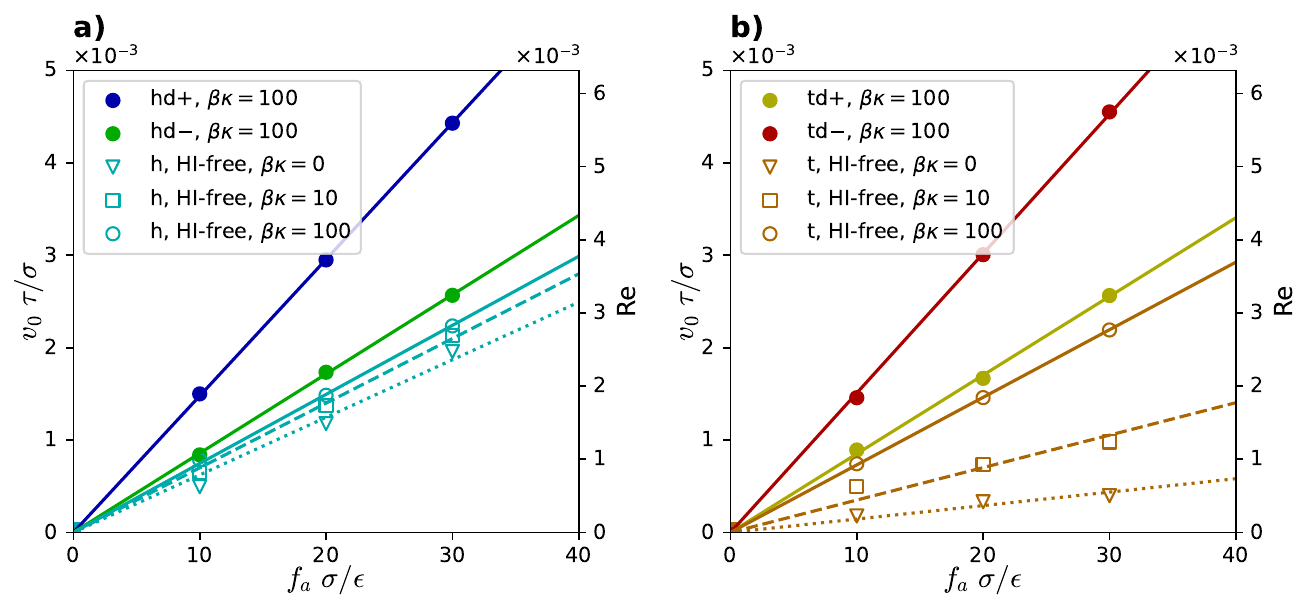}
    \caption{\label{fig:v0_fa_noHI}Swimming velocity $v_0$ against the active force $f_a$ for a) the head-active HI-free (h), b) the tail-active HI-free polymer (t), for varying stiffness parameters $\kappa$. We calculate $v_0$ as a time- and ensemble average over a time span of $\Delta t_{\mathrm{span}} = 9000\,\tau$ and 48 samples. Error bars are smaller than symbol size. The lines represent linear fits forced through (0,0).}
\end{figure*}

Without a counterforce volume, we can only distinguish between head-active and tail-active polymers. When we compare the swimming velocities with and without HI 
(Fig.~\ref{fig:v0_fa_noHI}), we find that the presence of HI can increase the swimming velocity significantly, provided that the counterforce volume is placed in front or at the back of the polymer, as given in the hd$+$ and td$-$ models, respectively. 
Placing the counterforce volume close to a monomer, as is the case for the hd$-$ and td$+$ models, slows the polymer down, resulting in a swimming speed within the order of magnitude of swimming speeds without hydrodynamic interactions. 

Interestingly, however, apart from the quantitative differences  of the swimming velocities, we observe very similar polymer properties regardless of whether HI are included or not: Head-active polymers are hardly sensitive to their stiffness parameter $\kappa$, because this type of activity induces an effective stiffness, therefore straightening the backbone and helping the polymer keep its orientation over time. For tail-active polymers on the other hand, stiffness plays a very dominant role, as flexibility allows for increased bending and therefore strong decorrelation of the polymer backbone orientation in time. These effects are mainly induced by the activity itself, and are only slightly influenced by the hydrodynamic self-interaction of the polymer. In particular, we observe the same cat's tail phenomenon without HI, leading to an oscillatory part in their auto-correlation function of the polymer backbone over time, and as a consequence the same sub-diffusive regime (not shown). Quantitatively, the decorrelation time of the polymer backbone is smaller without HI, as expected from previous studies of polymer relaxation times with and without HI (not shown).\cite{Liebetreu2020,Schneck2024} Static properties, such as the radius of gyration and the 
$\cos \theta$-distribution within the polymer, are hardly influenced by HI, as expected.


\section{\label{sec:monomer_resolved}Monomer-resolved properties} 

\begin{figure*}
    \includegraphics[width=\linewidth]{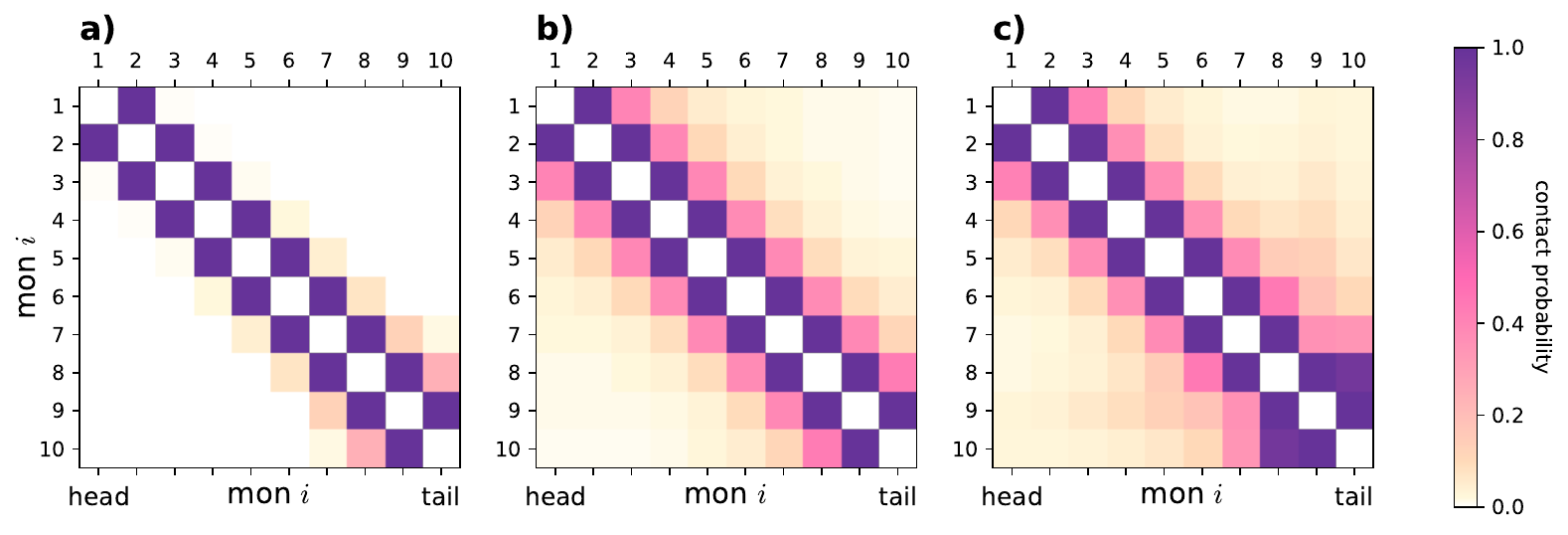}
    \caption{\label{fig:contacts}Contact probabilities of all monomer pairs in the fully flexible ($\beta\kappa=0$) polymer: a) hd$+$, $f_a=30\, \epsilon/\sigma$, b) passive, and c) td$-$, $f_a=30\, \epsilon/\sigma$. We define the monomer pair as "in contact" if it is separated by a distance $\leq 1.5\,\sigma$.}
\end{figure*}

In order to better understand the effects of the active force on the conformational properties of the polymer, we determine the contact maps for the fully flexible polymers, shown in Fig.~\ref{fig:contacts}.
The contact map displays the contact probability for each monomer pair as conformations change over time. We define a monomer pair to be "in contact" if it is separated by a distance $\leq 1.5 \sigma$. We observe that the fully flexible head-active polymer, Fig.~\ref{fig:contacts}a, exhibits far less overall contact between non-neighboring monomer pairs compared to the passive polymer,
shown in Fig.~\ref{fig:contacts}b, especially if one of the involved monomers is located near the active head monomer ($i=1$). This can be attributed to the effective stiffening effect caused by the head activity. On the other hand, the fully flexible tail-active polymer, Fig.~\ref{fig:contacts}c, shows overall more frequent contact between monomer pairs compared to its passive counterpart. We further find a significantly high contact probability between the tail monomer ($i=10$) and the monomer next to its neighbor ($i=8$). This demonstrates again the equilateral triangle formation caused by the active tail. The immediate neighbor of the active tail monomer ($i=9$) exhibits an increased contact probability with other monomers along the contour length. This can be attributed to the swirling motion in which the active tail monomer shoves its neighbor into the polymer core, causing it to spiral around like a cat's tail. In semi-flexible and stiff polymers, there is almost no contact between non-neighboring monomer pairs.

As previously discussed, to ensure a force-free system, we applied a 
counterforce on the solvent of equal magnitude and opposite direction as the force on the active monomer. The resulting forces on all of the individual monomers, however, are a consequence of redistributed forces stemming from internal interactions in the polymer and fluid flow.
%
%
To investigate how forces are redistributed, we calculate the 
total ensemble-averaged projection of the force 
${\boldsymbol{F}_{i}(t)}$ on  monomer $i$ 
along the polymer backbone $\hat{\boldsymbol{e}}_{\mathrm{ee}}(t)$,
namely:
\begin{equation}
    F_{i,||} = \left< {\boldsymbol{F}}_{i}(t) \cdot \hat{\boldsymbol{e}}_{\mathrm{ee}}(t) \right>.
    \label{eq:fparallel}
\end{equation}
%

\begin{figure*}
    \includegraphics[width=\linewidth]{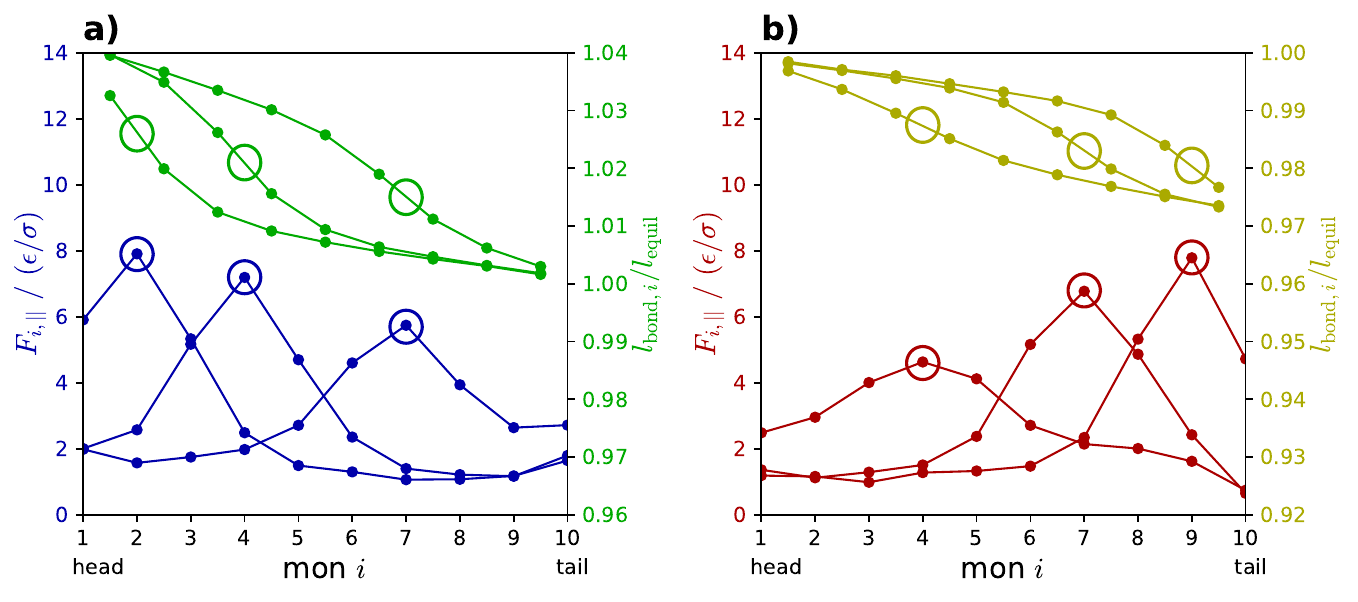}
    \caption{\label{fig:forces_per_mon}Left axes: The force $F_{i,||}$, 
    Eq.~(\ref{eq:fparallel}), acting on each monomer in the polymer, as it is redistributed from the applied active force on a) the head monomer, and b) the tail monomer, for varying displacements of the counterforce volume $d$, calculated as a time- and ensemble average over a time span of $\Delta t_{\mathrm{span}} = 9000\,\tau$ and 48 samples. Results are shown for the rigid polymers, $\beta \kappa = 100$, and a strength $f_a=30\,\epsilon/\sigma$ of the active force. The open circles indicate the location where the counterforce is applied. Right axes: The ratios of the local bond lengths $l_{\mathrm{bond},i}$ over the equilibrium bond length $l_{\mathrm{equil}}$.}
\end{figure*}

Fig.~\ref{fig:forces_per_mon} displays the force
$F_{i,||}$
along the polymer backbone for each individual monomer,
where the counterforce is applied behind the active head monomer (Fig.~\ref{fig:forces_per_mon}(a), hd$-$), i.e.\ around monomer $i=2$,
and in front of the active tail monomer (Fig.~\ref{fig:forces_per_mon}(b), td$+$), i.e., around monomer $i=9$, for the stiff polymers.
To better understand the effect of the specific counterforce location, we now vary the counterforce distance $d$ leading to different monomer positions where the counterforce is applied, depicted by the different curves in Fig.~\ref{fig:forces_per_mon}.
%
We note that, while $\sum \mathbf{F}_{i} = \mathbf{F}_a$ is always fulfilled,
the parallel components only fulfill the approximate relation $\sum F_{i,||} \approx f_a$ as a consequence of the
remaining flexibility even in a stiff polymer, and therefore the active force vector and the end-to-end vector of the polymer are not perfectly aligning. As can be seen in Fig.~\ref{fig:forces_per_mon}, the active force, that is applied to only either the head- or the tail monomer, is redistributed in a way such that the strongest force along the polymer backbone lies with the monomer that is closest to the counterforce volume, as is indicated by an open circle in Fig.~\ref{fig:forces_per_mon}. 
This is
due to the application of the active counterforce within the counterforce volume, which creates a strong local backflow in the fluid, as also visible in the measured flow fields (Fig.~\ref{fig:flow_fields}), resulting in
strong local drag forces balanced by the monomers.

By measuring the local bond lengths $l_{\mathrm{bond},i}$ across the polymer, normalized by the equilibrium bond length of the passive polymer $l_{\mathrm{equil}}$, we identify that in the head-active polymers all bonds are stretched, i.e., longer than the equilibrium bond length, whereas in the tail-active polymers all bonds are compressed (Fig.~\ref{fig:forces_per_mon}). The locations of the counterforce volumes, indicated by the open circles, align with the inflection points of the bond length curves. The inflection points of the bond length curves are another indicator to show where the local force is strongest, because where the bond length slope is at maximum, the monomer experiences the strongest force. Note also that in the case of the head-active monomer, all bonds 
are stretched, $l_{\mathrm{bond},i}/l_{\mathrm{equil}} > 1$, in comparison with the 
passive polymer, whereas the same become compressed, $l_{\mathrm{bond},i}/l_{\mathrm{equil}} < 1$, for the tail-active one.

\section{\label{sec:force_dipole}Force transmission and dipole strength} 

\begin{figure*}
    \includegraphics[width=\linewidth]{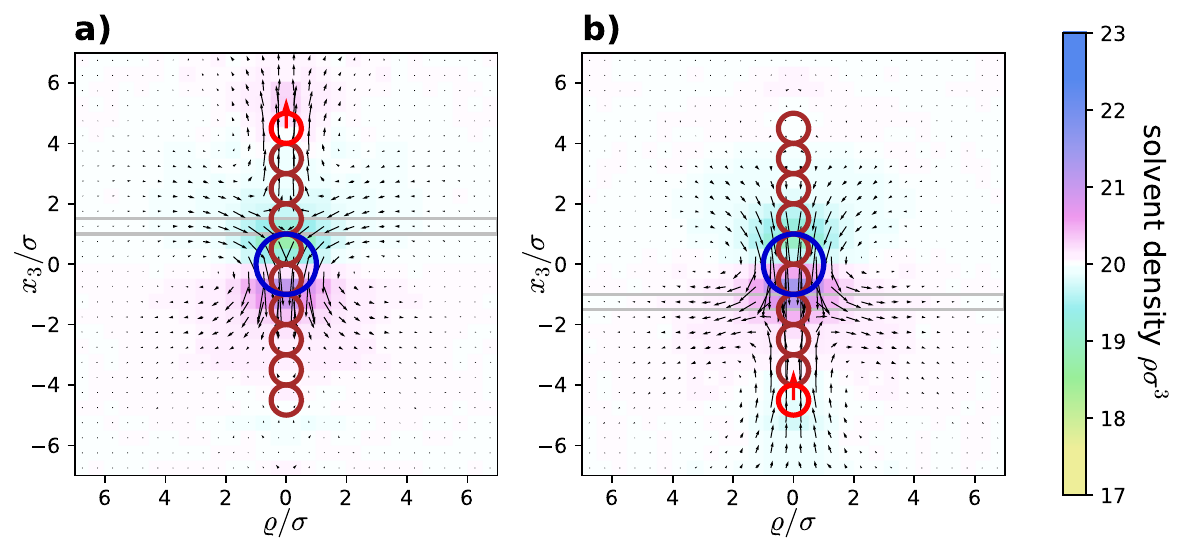}
    \caption{\label{fig:flow_center}Density- and flow fields of a) a head-active stiff polymer with $d=-4.5\,\sigma$, and b) a tail-active stiff polymer with $d=+4.5\,\sigma$ ($\beta\kappa=100$, $f_a=30\,\epsilon/\sigma$). The blue circle represents the counter force volume, red circles represent monomers. The active monomer and active force vector are depicted by the bright red circle and arrow respectively.
    }
\end{figure*}

\begin{figure*}
    \includegraphics[width=\linewidth]{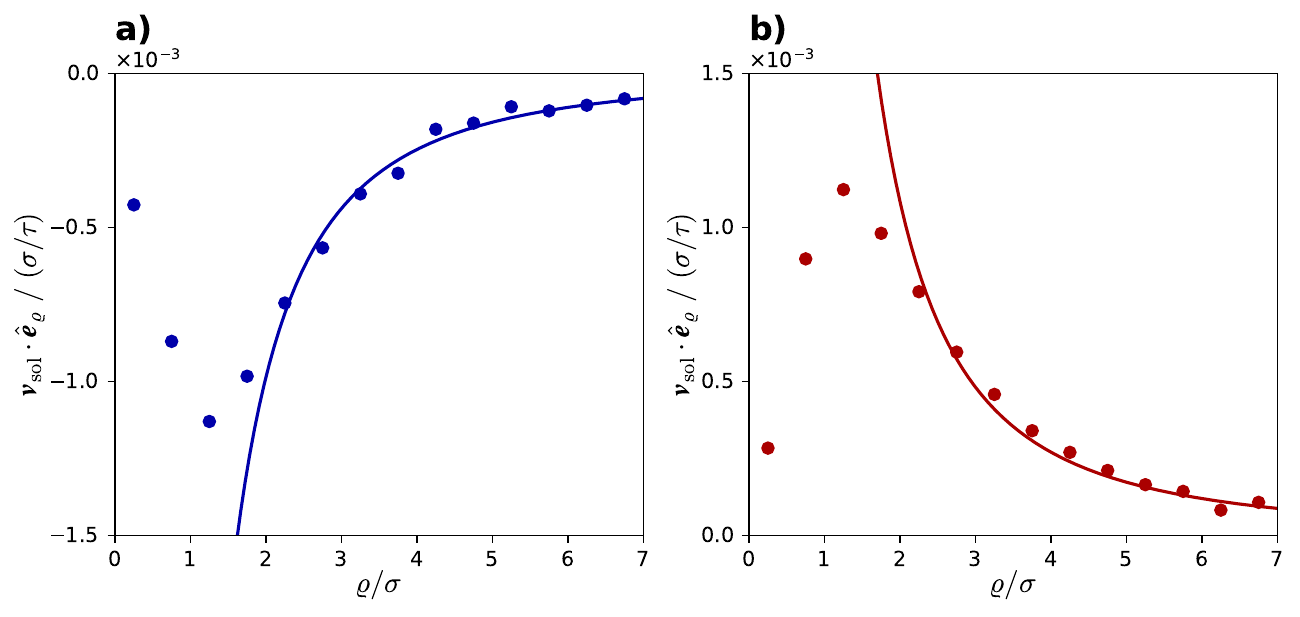}
    \caption{\label{fig:flow_field_decay}The solvent velocity along $\hat{\boldsymbol{e}}_{\varrho}$ decreases with distance perpendicular to the polymer backbone following $\propto \varrho^{-2}$ in the far field, as depicted for a) the head-active polymer with $d=-4.5\,\sigma$ and b) the tail-active polymer with $d=+4.5\,\sigma$ (both $\beta \kappa = 100$, $f_a=30\,\epsilon/\sigma$). Measured values of the solvent velocity along $\hat{\boldsymbol{e}}_{\varrho}$ are shown by the dots. The solid lines
    represent fits according to Eq.~(\ref{eq:dipole_strength}), with force dipole
    strengths $p = 15.7\,m\sigma^2/\tau^2$ for 
    panel a) and $p = -17.3\,m\sigma^2/\tau^2$ for
    panel b).}
\end{figure*}

Extrapolating from the thought that the active force is transmitted to other monomers in the chain, it would be reasonable to assume that the stiff head-active polymer with $d=-4.5\sigma$ exhibits the same flow field as the stiff tail-active polymer with $d=+4.5\sigma$. This 
assumption is based on the fact that for both the counterforce center aligns with the theoretical center-of-mass, and due to their stiff nature we would expect the force-dipole contribution to vanish and higher order hydrodynamic multipoles to become important.\cite{KimKarila}
However, as just discussed, even our stiff polymers are not perfectly rigid rods and forces are redistributed unevenly within the polymer, which affects the overall flow fields.
Fig.~\ref{fig:flow_center} shows that while the flow fields are very similar between the two models, as the active force that is applied to the active monomer gradually attenuates as it 
gets transmitted to the other monomers, the resulting flow fields slightly differ.
%
In fact, we establish that for these cases the far-field flow is still a force dipole, either a pusher or a puller, for $d=-4.5\sigma$ and $d=+4.5\sigma$, respectively. 
Force dipoles create flow fields that decay with the distance to the dipole-center in perpendicular direction $\varrho$ in the far field following \cite{Zottl2016}
\begin{equation}
    \boldsymbol{v}_{\mathrm{sol}} \cdot \hat{\boldsymbol{e}}_{\varrho} = -\frac{p}{8\pi\eta} \varrho^{-2},
    \label{eq:dipole_strength}
\end{equation}
with $p$ being the force dipole strength, where $p>0$ for pushers, and $p<0$ for pullers. In Fig.~\ref{fig:flow_field_decay}, we measure the local solvent velocities along $\hat{\boldsymbol{e}}_{\varrho}$ at the $x_3$-value where they are at maximum, indicated by the gray box in Fig.~\ref{fig:flow_center}, and then fit the curve to follow Eq.~(\ref{eq:dipole_strength}) in the far field.
As can be seen, the flow fields can be well approximated by force-dipoles of comparable strength but being either pusher or puller,
where $p=15.7\,m\sigma^2/\tau^2$ (pusher) for $d=-4.5\,\sigma$ and $p=-17.3\,m\sigma^2/\tau^2$ (puller) for $d=+4.5\,\sigma$.

\section{\label{sec:conclusions}Conclusions and outlook} 

In this work, we have introduced a new hydrodynamic model for active polymers where the hydrodynamic flow fields around the polymers can be tuned by the position of the active force and counterforce applied locally to the surrounding fluid.
We have shown that the swimming velocity of an active stiff polymer increases linearly with the applied active force in a low Reynolds number environment, as expected. However, for active flexible polymers, the relationship is nonlinear, and depends on the location of the active site. Additionally, the backbone orientation auto-decorrelates faster for tail-active compared to head-active polymers, and faster for flexible compared to stiff polymers. The dynamics of sufficiently stiff active polymers are very similar to active rods, i.e., they follow the mean-squared-displacement predictions from the persistent random walk model, whereas tail-active flexible polymers exhibit sub-diffusive regimes due to a circulative motion that resembles that of a cat's tail. While angular momentum is not perfectly conserved in our system, it fluctuates around zero without building up any net angular momentum.

The activity of the monomer influences the polymer conformation, and depending on the location of the active site the polymer adopts conformations resembling typical conformations of stiffer or less stiff versions of passive polymers. Fully flexible tail-active polymers frequently adopt an equilateral triangle formation between their last three monomers. The conformational effects that the active monomer causes, however, decrease along the polymer backbone the further we move away from the active site.

Simulations featuring the MPCD-solvent include hydrodynamic interactions, and the counterforce we apply to the solvent in order to simulate a force-free system can be located at different positions relative to the polymer. Depending on where the counterforce is applied, the active polymer creates different kinds of flow fields, some of which resemble pusher and puller dipole flow fields, but are in general described by higher order multipole flow fields. The location of the counterforce also influences the swimming velocity, making the polymer a more efficient swimmer if the counterforce is applied further away from the polymer.

Switching hydrodynamics off by using the random MPCD solvent instead, we observed a slower swimming speed of the active polymer compared to the hydrodynamic case, as well as a faster auto-decorrelation of the polymer backbone orientation. However, hydrodynamic interactions within a polymer did not significantly influence conformational properties.
That said, we expect that in more concentrated solutions of active polymers, which we plan to investigate in the future, hydrodynamic effects are expected to play an important role, as known in previous hydrodynamic models of active rods, ellipsoids and spheres.\cite{Saintillan2007,Zottl2014,Zantop2022}

\section*{Supplementary Material}
See the supplementary material for videos of the active polymer simulations.

\begin{acknowledgments}
This work has been supported 
by the European Union through the Twinning project FORGREENSOFT (Grant No.~101078989 under HORIZON-WIDERA-2021-ACCESS-03).
The computational results have been achieved using the Austrian Scientific Computing (ASC) infrastructure.

A CC-BY public copyright license has been applied by the authors to the present document and will be applied to all subsequent versions up to the Author Accepted Manuscript (alternatively final peer-reviewed manuscript accepted for publication) arising from this submission, in accordance with the grant’s open access conditions.
\end{acknowledgments}

\section*{Author Declaration}
\subsection*{Conflict of Interest}
The authors have no conflicts to disclose.
\subsection*{Author Contributions}
\textbf{Lisa Sappl:} Performing the simulations (lead); Data curation (lead); formal analysis (lead); methodology (supporting); software (lead); visualization (lead); writing – original draft (lead); writing – review and editing (equal). \textbf{Christos N. Likos:} Conceptualization (supporting); formal analysis (supporting); funding acquisition (lead); methodology (equal); project administration (equal); resources (lead); supervision (equal); validation (equal); writing – review and editing (equal). \textbf{Andreas Z\"ottl:} Conceptualization (lead); formal analysis (supporting); methodology (equal); project administration (equal); resources (supporting); supervision (equal); validation (equal); writing – original draft (supporting); writing – review and editing (equal).

\section*{Data Availability}
The data that support the findings of this study are available from the corresponding author upon reasonable request.

\appendix*

\section{\label{sec:dimers}Active dimers} 

In order to access all relevant time scales, we hereby show simulation results using dimers, i.e., polymers that only consist of $N=2$ monomers. Due to their smaller size, the backbone orientation of dimers auto-decorrelates up to two orders of magnitude faster than with decamers ($N=10$). Further, the necessary simulation box size is also a lot smaller, we operate at $L=8\,\sigma$, both making simulations of longer time scales more feasible. A notion of stiffness is not sensible for dimers, since there need to be three monomers to form an angle between two bonds. As such, dimers are perfectly stiff.

\begin{figure}
    \includegraphics[width=\linewidth]{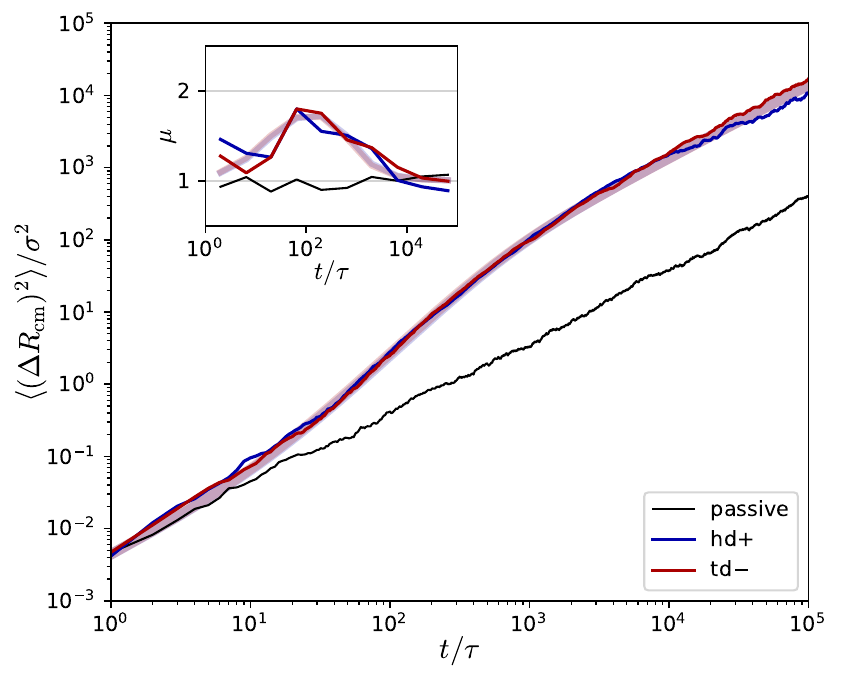}
    \caption{\label{fig:msd_dimer}Mean squared displacement over time for the active dimers hd$+$ (blue) and td$-$ (red) compared to a passive dimer (black), $f_a=30\,\epsilon/\sigma$. The inset shows the exponent $\mu$ of the scaling in $\langle (\Delta R_{\mathrm{cm}})^2 \rangle \propto t^{\mu}$ in the corresponding mean-squared displacement curves. The bright bold curves represent the theoretical MSD and its exponent (inset) of persistent random walks for stiff particles following Eq.~(\ref{eq:persistent_rw}).}
\end{figure}

The mean-squared displacement (MSD) of a hd$+$ dimer and a td$-$ dimer over time is shown by Fig.~\ref{fig:msd_dimer}.
The time scales are long enough for the dimer to enter the second diffusive regime, which will persist from there on. We can see that 
the MSD-plots for both models agree well with each other, and both follow the theoretical persistent random walk model closely.


\providecommand{\noopsort}[1]{}\providecommand{\singleletter}[1]{#1}%

\end{document}